\documentclass[12pt,a4paper]{article}
\usepackage{placeins}
\usepackage[title]{appendix}

\usepackage{xcolor}

\usepackage[british]{babel}
\usepackage[a4paper,top=2cm,bottom=2cm,left=2.5cm,right=2.5cm,marginparwidth=1.75cm]{geometry}

\usepackage[utf8]{inputenc}
\usepackage[T1]{fontenc}
\usepackage{lmodern}
\usepackage[scaled=0.95]{helvet}

\usepackage{setspace}
\usepackage{titlesec}
\titleformat{\section}{\normalfont\Large\bfseries}{\thesection.}{1em}{}

\usepackage{caption}
\usepackage{subcaption}
\usepackage{amsmath,amsfonts}

\usepackage{csquotes}
\usepackage[style=apa,backend=biber]{biblatex}
\DeclareLanguageMapping{british}{british-apa}
\DeclareFieldFormat[article]{volume}{\apanum{#1}}

\usepackage{graphicx}
\usepackage{booktabs}
\usepackage{threeparttable}
\usepackage{diagbox}
\usepackage{authblk}
\usepackage{tikz}
\usetikzlibrary{positioning,shapes,arrows.meta}
\usepackage{enumitem}
\usepackage{url}
\usepackage{comment}
\usepackage{hyperref}
\hypersetup{hidelinks}
\usepackage{float}
\usepackage{wrapfig}
\usepackage[normalem]{ulem}

\begin{document}




    
    



    
    









\clearpage

\begin{center}

{\LARGE Physics-Guided Multi-Objective Deep Learning for Ultrasound RF Data Interpolation in Resource-Constrained Imaging}\\[3ex]

{\large
Luoyuan Zhang$^{1}$,
Yiyang You$^{2}$,
Ananya Tandri$^{2}$,
Yinan Feng$^{1}$,\\
Hyunwoo Song$^{2}$,
Jeeun Kang$^{2,~\ast}$, and
Youzuo Lin$^{1,~\ast}$
}\\[1.2em]

{\small
$^{\textbf{1}}$University of North Carolina at Chapel Hill, Chapel Hill, NC, USA \\
$^{\textbf{2}}$Johns Hopkins University, Baltimore, MD, USA \\
$^{\boldsymbol{\ast}}$Co-corresponding authors:
\texttt{jkang@jhmi.edu} and \texttt{yzlin@unc.edu}
}
\end{center}

\begin{abstract}

Ultrasound imaging increasingly targets portable, point-of-care, and wearable settings where constraints on power, bandwidth, and hardware complexity often necessitate sparse data acquisition in spatiotemporal scanning.
However, image reconstruction using the sparse data can introduce insufficient phase information in coherent beamforming process, resulting in grating-lobe artifacts that degrade imaging contrast resolution.
We present a physics-guided, data-driven framework for sparse-to-dense radio-frequency (RF) reconstruction that aligns training with downstream image formation.
Our approach trains an end-to-end interpolation network using a hybrid supervision scheme that combines an RF-domain and a beamforming-domain loss with exponential moving average (EMA) to stabilize the multi-objective training.
To improve generalization under variable acquisition layouts, we also introduce a random-skip masking strategy that varies sparsity patterns during training so a single model can handle diverse decimation factors and irregular channel configurations.
We evaluate the framework on a held-out test set using the mean structural similarity index measure (SSIM) between reconstructed and ground-truth beamformed images.
Across decimation factors $\times2$ to $\times13$, the best-performing configuration maintains mean SSIM around 0.95.
Overall, the results show consistent gains in RF reconstruction and post-beamforming image quality across diverse acquisition conditions.
This approach enables robust, high-quality ultrasound imaging at resource-constrained settings by allowing more sparse scanning in spatiotemporal domain.

\end{abstract}


\section{Introduction}

Ultrasound is rapidly moving beyond conventional cart-based scanners toward point-of-care, portable, and increasingly wearable or hands-free devices \parencite{zhou2025wearable, kim2012single, kang2015system}, enabling continuous and operator-independent imaging and monitoring of lesions in intensive care settings.
Clinical efficacy of these emerging platforms is typically constrained by necessary spatial field-of-view (FOV) and temporal resolution to capture spatiotemporal dynamics in the target lesions, power budget, on-device computation, and (often wireless) data transfer bandwidth, which limit the access to full spatiotemporal sampling resolutions. \parencite{huang2024wearable,lei2025wearable}.
Straightforward solutions may consider sparse array or time-multiplexed spatial sampling; however, such sub-sampling effectively increases inter-element spacing and can introduce strong grating-lobe artifacts, degrading imaging contrast resolution \parencite{kim2019gapfilling,qi2023sparsearray,tanter2014ultrafast}.
Motivated by this tension between hardware constraints and image quality, there is growing interest in reconstructing or synthesizing dense radio-frequency (RF) or channel data from sparse measurements \parencite{yoon2019efficient}.

Early efforts primarily relied on classical signal-processing approaches to compensate for missing channels.
Representative examples include linear or spline interpolation along the sparsely sampled channel data, low-rank and annihilating-filter based reconstruction, and related matrix-completion style formulations
\parencite{deboor2001splines,jin2016aloha_us,candes2009matrixcompletion}.
While these methods can work under mild undersampling, they often depend on smoothness or low-rank assumptions and may degrade at high decimation, leaving phase-incoherent errors that amplify after coherent beamforming \parencite{kim2019gapfilling}.

More recently, deep learning has been explored for ultrasound reconstruction and image formation from channel data.
One line of work focuses on RF-domain interpolation, using encoder--decoder architectures such as U-Net to reconstruct dense channel measurements from sparse inputs and often outperforming classical interpolation baselines
\parencite{xiao2022minimizing,MamistvalovDL}.
A complementary direction targets image formation more directly, including learned beamforming and networks designed to suppress clutter and off-axis scattering or to approximate adaptive beamformers, which can improve imaging quality compared with conventional delay-and-sum (DAS) under sparse sampling scenarios \parencite{luchies2018dnn,nair2018altbf,hyun2019speckle,khan2020adaptive}.
Other studies leverage image-domain supervision for limited-acquisition imaging, such as enhancement for plane-wave imaging, self-supervised reconstruction from plane-wave RF data, and generative adversarial network (GAN)-based mappings from sparse measurements to higher-quality beamformed images \parencite{qi2021pw,zhang2021selfsup,zhou2021gan}.

Despite these advances, two practical issues remain.
First, many RF interpolation methods are trained to minimize RF-domain errors only (e.g., mean squared error between predicted and dense RF) \parencite{yoon2019efficient}. 
This can introduce an \emph{objective mismatch}: small phase and/or amplitude errors that appear negligible under an RF-domain loss may coherently accumulate during DAS beamforming, manifesting as elevated grating-lobes and degraded contrast resolution in the final image \parencite{luchies2018dnn,khan2020adaptive}. 
Second, both RF-domain interpolation and image-domain enhancement pipelines are often trained and evaluated under \emph{fixed} sub-sampling or acquisition patterns (e.g., pre-defined decimation layouts or fixed transmit settings), which can limit robustness when channel layouts are unknown, variable, or time-varying in practical scenarios such as wearable or freehand probes \parencite{yoon2019efficient,hyun2021cubdl}.
In such settings, these methods may reduce visible artifacts under a specific acquisition regime but can still exhibit structured residual artifacts when sparsity or layouts change.
Together, objective mismatch and fixed-pattern training can make sparse-reconstruction pipelines brittle under real-world acquisition variability.

A natural way to address the objective mismatch is to incorporate the image-formation operator into learning.
Along this line, physics-informed machine learning embeds known forward-model structure into training, often by using differentiable operators and defining losses on physically meaningful outputs rather than only on intermediate signals \parencite{lin2024cwi_survey,raissi2019pinn,adler2018learnedprimaldual,aggarwal2018modl}.
This perspective is particularly relevant for ultrasound, since coherent beamforming is highly sensitive to phase and amplitude consistency in RF signals.

To follow this physics-guided learning principle and address the limitations of prior deep learning pipelines, we propose a physics-guided sparse-to-dense RF reconstruction framework that aligns learning with the downstream imaging objective. 
Our method trains an interpolation network with hybrid supervision that couples RF-domain fidelity as an input objective with a beamforming-domain output objective to resemble the ground-truth beamformed results. By doing so, the combined loss function better aligns optimization with the final imaging objective.
We further stabilize this multi-objective training with exponential moving average (EMA)-based adaptive loss balancing \parencite{polyak1992averaging} and improve generalization by randomizing sparsity patterns during training, enabling a single model to operate robustly under any, even out-of-distribution (OOD) sampling layouts.

In this paper, we instantiate the interpolation network using two representative encoder backbones: HGNet-V2 and CAFormer \parencite{Yu2024MetaFormerBaselines}. Both variants are trained using the same hybrid RF- and beamforming-domain supervision, EMA-based adaptive loss weighting, and random-skip masking strategy.

The remainder of this paper is organized as follows. Section~2 introduces the relevant background and formulates the sparse RF reconstruction problem, including DAS beamforming and grating-lobe formation under sparse sampling. Section~3 presents the proposed physics-guided learning framework, including the hybrid training objective, EMA-based adaptive loss weighting, and random-skip masking strategy. Section~4 describes the network architecture and the two encoder-backbone variants. Section~5 presents the simulation dataset and training and evaluation setup. Section~6 reports the experimental results, including RF-domain reconstruction, grating-lobe suppression, image-level reconstruction quality, robustness to varying sparsity and noise, ablation studies, and preliminary generalization experiments on a synthetic resolution phantom. Finally, Section~7 summarizes the main findings, discusses the limitations of the present study, and outlines directions for future work.

\section{Background and Problem Formulation}
\label{sec:background}
In this section, we introduce the background and problem formulation for sparse-to-dense ultrasound RF reconstruction. We first review DAS beamforming and establish notation for the beamforming operator. We then describe  the sparse acquisition setting, the reconstruction task, and a standard RF-domain training objective that serves as a baseline for the physics-guided framework presented in the next section.

\subsection{DAS Beamforming Preliminaries}
Ultrasound imaging transmits an acoustic pulse and records backscattered echoes on a receive array. Under a homogeneous sound-speed model, the RF waveform measured at element $e$, denoted by $v_e(t)$, can be viewed as a superposition of delayed and weighted signals from reflectors in spatial grid $\mathbf{r}$, where delays are governed by acoustic time-of-flight to each channel of the ultrasound imaging array. Image formation aims to coherently combine multi-channel measurements so that echoes from a chosen focal location add constructively while off-axis energy is suppressed.

DAS is a standard image-formation approach, repeating this process for individual imaging pixels in the target imaging frame. Let $\tau_e(\mathbf{r})$ denote the expected two-way delay for echoes originating from location $\mathbf{r}$ to be observed at receive element $e$. DAS evaluates each channel at its corresponding delay and sums across the aperture:
\begin{equation}
x(\mathbf{r}) = \sum_{e=1}^{S} w_e(\mathbf{r}) \, v_e\!\big(\tau_e(\mathbf{r})\big),
\end{equation}
where $w_e(\mathbf{r})$ denotes an apodization weight (and may incorporate dynamic aperture selection) to control sidelobes. $\tau_e(\mathbf{r})$ compensates different time-of-flight in $v_e(t)$ over the imaging aperture, and their summation will restore  the initial pressure from the pixel. A beamformed data $x(\mathbf{r})$ is finally obtained by repeating the process over $\mathbf{r}$, a grid of depths and scanlines, followed by envelope detection and log compression for clinical readability.

\subsection{Grating Lobe Artifacts and Suppression}
\label{sec:problem_setup}
Let $V_{\text{full}} \in \mathbb{R}^{D \times S}$ denote fully sampled receive RF data at spatial sampling rate enough to suppress grating lobe artifacts. Here, $D$ is the number of temporal~(depth) samples and $S$ is the number of sensing events that configure the full aperture. Under sparse acquisition, only a subset of channels is observed, resulting in a sparse input denoted by $V_{\text{sparse}}$. Its explicit mask-based construction is described in Section~\ref{sec:random}.


The image reconstruction using $V_{\text{sparse}}$ will result in grating lobe artifacts where limited spatial sampling produces false constructive interferences by failing to obtain enough phase information to destruct the off-axis signals \parencite{paul1997side}. In analytical form, it is well known that a beam profile at the focal point can be indicated as the Fourier transform of the spatial sampling function \parencite{song2022novel, song2023synthetic, yoon2022elevational}. For simple example, the beam profile from a linear aperture would be in sinc function, defined by the overall aperture, which repeats at the spatial sampling frequency (grating lobes). The centricity to the imaging aperture will be applied with a window defined by sub-aperture function defined in each sensing event. In clinically available ultrasound probes, manufacturers design the piezoelectric array to keep the grating lobes sufficiently suppressed with the small element pitch. \textbf{$V_{\text{sparse}}$} with lower spatial sampling frequency would bring the grating lobes closer into the main beam with higher intensity. 
During radial or elevational scanning, sparse acquisition increases the effective spacing between successive scan planes, which can introduce grating-lobe artifacts after beamforming.

Given $V_{\text{sparse}}$, the goal is to recover a dense estimate $\hat V \in \mathbb{R}^{D \times S}$ that approximates $V_{\text{full}}$ to suppress the physics-defined grating lobe artifacts. We learn an interpolator $I(\cdot;\theta)$ that maps sparse inputs to dense outputs
\begin{equation}
\hat{V} = I(V_{\text{sparse}}; \theta),
\end{equation}
where $\theta$ denotes the trainable parameters of the interpolation network. A standard baseline trains $I(\cdot;\theta)$ with an RF-domain reconstruction loss
\begin{equation}
\mathcal{L}_{\text{RF}}(\theta) = \| \hat{V} - V_{\text{full}} \|_1,
\end{equation}
where $\|\cdot\|_1$ denotes the element-wise $\ell_1$ norm, i.e., the sum of absolute differences over all entries. Although RF-domain supervision promotes sample-wise agreement with the dense target, projected similarity metric in RF channel data does not explicitly capture how it affects coherent beamforming performance in Eq.~(1). This motivates physics-guided multi-objectives that align sparse-to-dense RF reconstruction with downstream image formation, introduced in the next section.

\section{Physics-Guided Learning Framework}
\label{sec:method}

\begin{figure}[!ht]
  \centering
  \includegraphics[width=\linewidth]{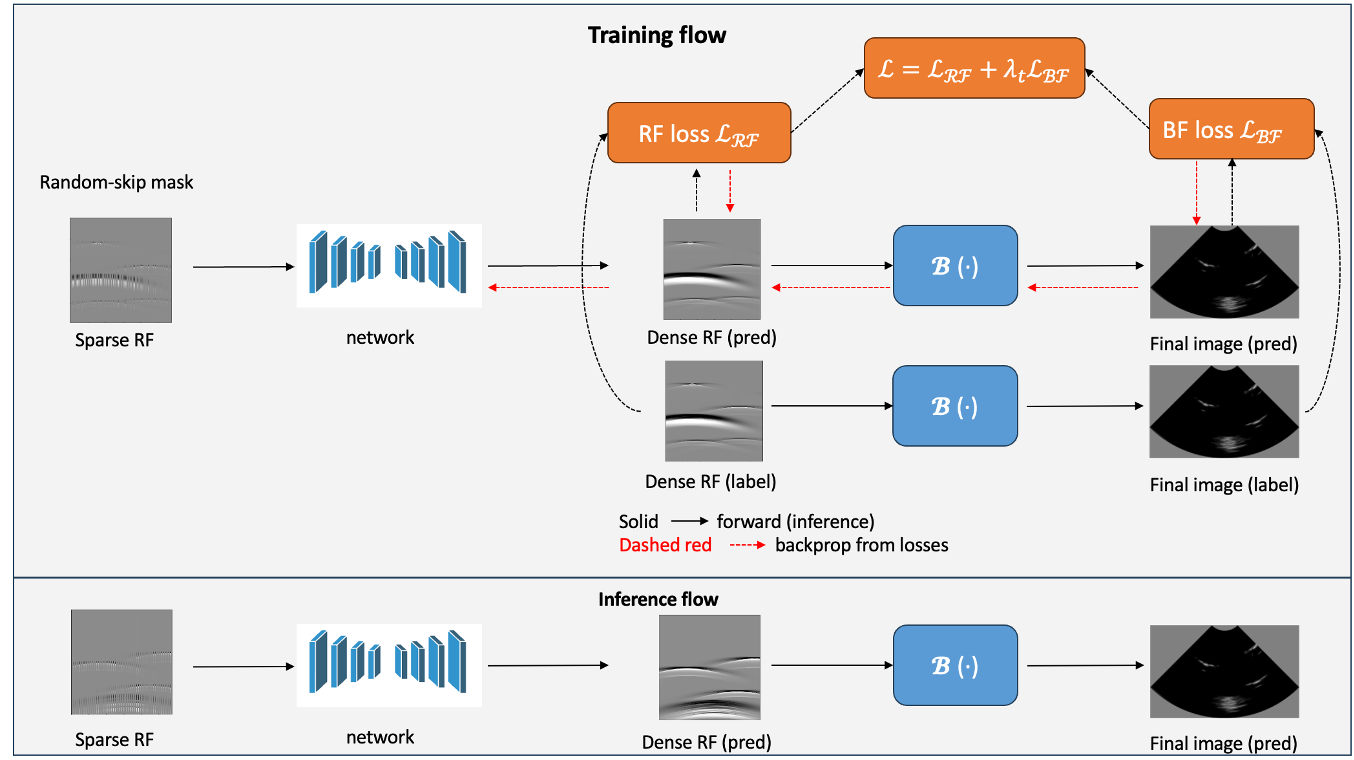}
  \caption{Overview of our physics-guided pipeline for training and inference. 
A random-skip mask produces the sparse RF input, which a neural network maps to a dense RF estimate. 
Training is supervised by an RF \(\ell_1\) loss and an image-domain MSE computed from the beamformed images; 
the beamforming loss is weighted by \(\lambda_t\), updated via an Exponential Moving Average (EMA). 
Red dashed arrows indicate backpropagation paths. 
Bottom: inference path from sparse RF \(\rightarrow\) neural network \(\rightarrow\) dense RF \(\rightarrow\) \(B(\cdot)\) \(\rightarrow\) final image.}
  \label{fig:teaser}
\end{figure}

In this section, we present our physics-guided learning framework for sparse-to-dense ultrasound RF reconstruction. The key idea is to train an interpolation network with hybrid supervision that couples RF-domain fidelity with downstream beamforming quality, while improving robustness to variable acquisition layouts through randomized sparsity patterns. Figure~\ref{fig:teaser} provides an overview of the training and inference flows and illustrates how the interpolation network interacts with the beamformer within the learning loop. 
We first introduce the hybrid objective that combines an RF-domain loss with a beamforming-domain loss. We then describe EMA-based adaptive loss balancing and the dynamic random-skip masking strategy, and finally detail the beamforming module that enables scalable end-to-end optimization.

\subsection{Physics-Guided Objective}
\label{sec:physics_guided_objective}

As introduced in Section~\ref{sec:problem_setup}, an interpolator $I(\cdot;\theta)$ can be trained with an RF-domain loss $\mathcal{L}_{\text{RF}}$ to recover a dense estimate $\hat V$ from sparse input $V_{\text{sparse}}$. However, RF-only supervision can be misaligned with the imaging objective because small amplitude or phase inconsistencies that appear minor in the RF domain can accumulate and manifest as ineffective suppression of grating lobe artifacts and low contrast resolution in the beamformed data.

To better align learning with image formation, we introduce a beamformer $B(\cdot)$ that discretizes the DAS operator in Eq.~(1) on a fixed grid of target imaging pixels.
We then supervise the network in the beamforming domain by matching the beamformed outputs of the prediction and the ground-truth:
\begin{equation}
\mathcal{L}_{\text{BF}}(\theta) = \| B(\hat{V}) - B(V_{\text{full}}) \|_2^2,
\end{equation}
where $\|\cdot\|_2^2$ denotes the squared Euclidean norm over all entries of the beamformed data.
The resulting hybrid objective function is
\begin{equation}
\mathcal{L}(\theta) = \mathcal{L}_{\text{RF}}(\theta) + \lambda \, \mathcal{L}_{\text{BF}}(\theta)
= \| \hat{V} - V_{\text{full}} \|_1 + \lambda \| B(\hat{V}) - B(V_{\text{full}}) \|_2^2,
\label{eq:totalloss}
\end{equation}
where $\lambda$ is a weighting coefficient that balances the contributions of RF- and image-domain supervision. This hybrid formulation encourages the network to recover dense RF data that not only maximizes similarity to the target RF signals but also yields structurally accurate, artifact-suppressed beamformed outcome. To keep both terms comparably influential during training, we use an adaptive weighting that updates $\lambda$ based on the running magnitudes of $\mathcal{L}_{\text{RF}}$ and $\mathcal{L}_{\text{BF}}$ (Section~\ref{sec:adaptive}). Embedding the beamformer in the loss directly optimizes post-beamforming quality by penalizing errors that are most visible after image formation.

\subsection{Adaptive Weighting via Exponential Moving Average}
\label{sec:adaptive}
Our total objective function in Eq.~\eqref{eq:totalloss} combines an RF-domain loss $\mathcal{L}_{\text{RF}}$ and a beamforming-domain loss $\mathcal{L}_{\text{BF}}$ with a balancing weight $\lambda$.
In practice, $\mathcal{L}_{\text{BF}}$ often decreases faster than $\mathcal{L}_{\text{RF}}$ during training.
With a fixed $\lambda$, this can gradually reduce the relative influence of the beamforming constraint and weaken the intended physics-guided supervision.
To address this issue, we replace the constant weight with an adaptive coefficient $\lambda_t$ that evolves over training to maintain a more consistent balance between the two objectives.

Because per-batch losses can fluctuate under stochastic optimization, we use an EMA to obtain a stable estimate of the target weighting coefficient.
Specifically, we first compute the instantaneous ratio
\begin{equation}
x_t = k \cdot \frac{\mathcal{L}_{\text{RF},t}}{\mathcal{L}_{\text{BF},t} + \varepsilon},
\end{equation}
where $k$ sets the desired relative scale and $\varepsilon$ is a small constant for numerical stability.
We then update the weight using EMA smoothing
\begin{equation}
\lambda_t = \beta \lambda_{t-1} + (1-\beta) x_t, \quad \beta \in (0,1),
\end{equation}
where larger $\beta$ yields smoother but slower updates, while smaller $\beta$ adapts faster but with higher variance.
In all experiments, we fix $\beta=0.9$, which provides a stable yet responsive estimate in practice.
This adaptive scheme allows $\lambda_t$ to change gradually during training, preventing either loss term from dominating and improving optimization stability across varying sparsity conditions.

\subsection{Dynamic Random-Skip Sampling for Generalization}
\label{sec:random}

\begin{figure}[!htbp]
  \centering
  \includegraphics[width=0.92\linewidth]{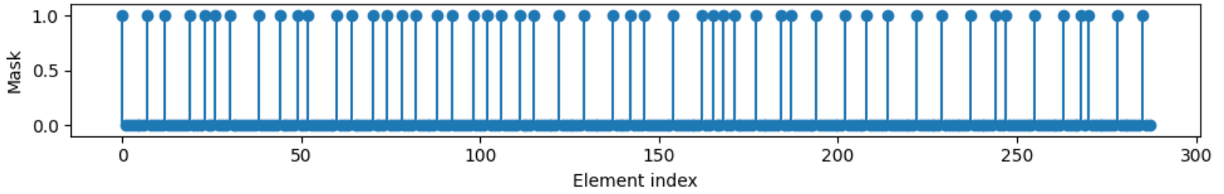}
  \caption{Illustration of the random-skip masking strategy (maximum skip length $R$ = 7). 
  Blue stems mark retained channels (value 1); gaps down to 0 indicate consecutively
  skipped channels. A new mask is sampled for each training instance and epoch,
  producing diverse, spatially varying sparsity patterns while keeping the input
  size $S$ unchanged.}
  \label{fig:random_skip_mask}
\end{figure}

Training is fully supervised using paired dense and sparse RF data. For each training example $i$, the sparse input is generated by applying a binary channel mask $M^{(i)} \in \{0,1\}^{S}$ to the dense RF frame

\begin{equation}
V_{\text{sparse}}^{(i)}(d,s)
=
M_{s}^{(i)}\, V_{\text{full}}^{(i)}(d,s),
\qquad
d=1,\ldots,D,\; s=1,\ldots,S,
\end{equation}
where $M^{(i)}_s=1$ retains the $s$th channel and $M^{(i)}_s=0$ masks it out.

Instead of using one fixed decimation layout for all samples, we randomize the mask during training. We generate $M^{(i)}$ using a random-skip rule controlled by a maximum skip length $R$. Starting from a retained channel, we skip a random number of consecutive channels, uniformly sampled between 1 and $R$, then retain the next channel, and repeat until the aperture is covered. Figure~\ref{fig:random_skip_mask} shows an example mask with $R=7$, where retained channels appear as stems and skipped runs appear as gaps. This procedure produces spatially varying effective decimation while keeping the number of sampling events $S$ unchanged.

This training-time randomization has two practical benefits. First, it provides an engineering advantage: sparsity is encoded by the mask within a fixed-length input, so the network input dimension stays the same across decimation factors and the same model can be deployed without modifying the architecture. Second, it improves generalization by exposing the interpolator to a wide range of missing-channel layouts during training, which makes performance more reliable when test-time sampling rates or channel configurations differ from those seen in training. We report an ablation study on the maximum skip length $R$ in Section~\ref{sec:rand-skip-ablation}.

\subsection{Efficient DAS Beamforming for Scalable Training}

A direct DAS implementation using sequential programming is slow when evaluated scanline by scanline. To make end-to-end training practical under the physics-guided hybrid objective, we implement the beamforming operator $B(\cdot)$ as a generic module with maximal vectorization and parallelization using graphics processing units (GPUs) for faster training/validation workflow.

Our implementation computes beamforming on any given full image grids in parallel, vectorized over target depth and scanlines. Geometry terms that are fixed for a given acquisition setup are precomputed once and reused during training. For each forward pass, we evaluate delays and apodization on the grid with batched tensor operations, and accumulate the delayed, weighted contributions across the receive-channel dimension to form the beamformed output.

This design avoids explicit loops over depth and output scanlines, providing an efficient implementation of the beamforming operator during training. Since the beamforming operator is differentiable, gradients from the beamforming-domain loss can backpropagate through $B(\cdot)$ to the interpolation network, enabling direct optimization for post-beamforming image quality.
\section{Network Architecture}
\label{sec:architecture}

\subsection{Backbone Design and Selection}
\label{sec:backbone}

\begin{figure}[!ht]
  \centering
  \includegraphics[width=\linewidth]{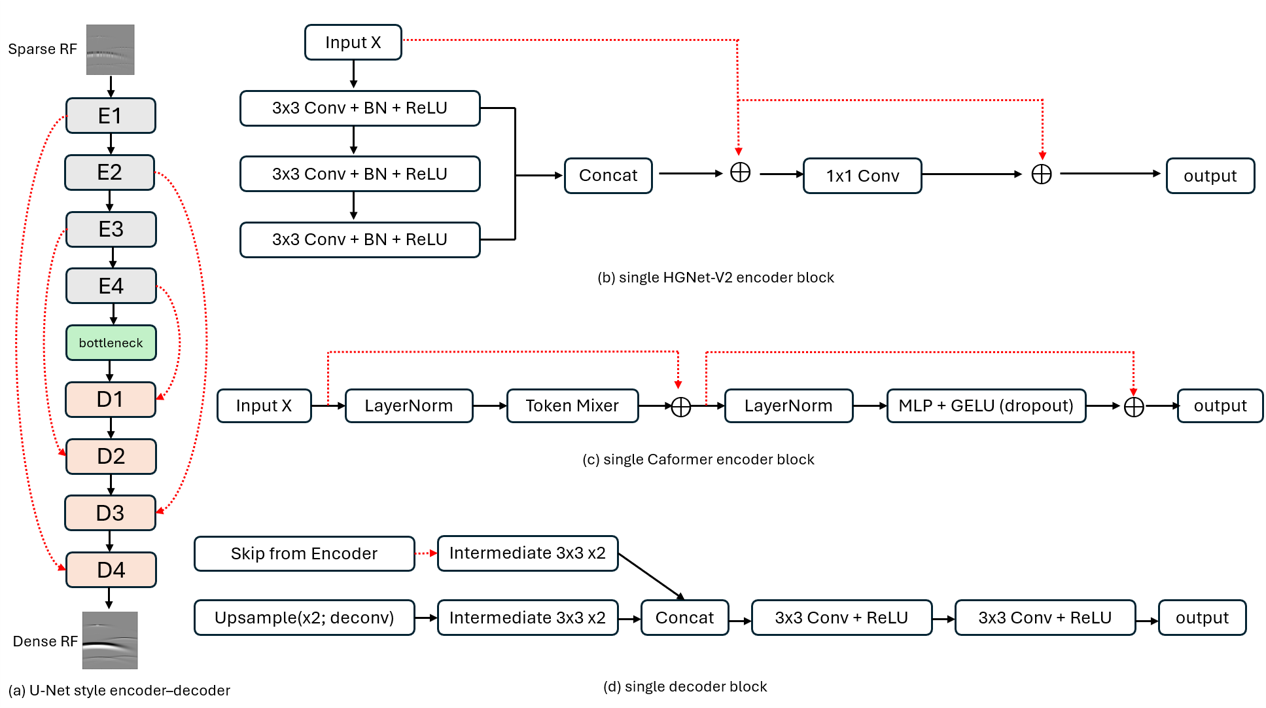}
  \caption{Architecture and block design.
  (a) U\mbox{-}Net style encoder--decoder with four encoder stages (E1--E4), a bottleneck, and four decoder stages (D1--D4).
  (b) HGNet-V2 encoder block: three parallel $3{\times}3$ Conv--BN--ReLU branches, concatenation, and a $1{\times}1$ projection with a residual connection.
  (c) CAFormer encoder block: LayerNorm $\rightarrow$ token mixer $\rightarrow$ LayerNorm $\rightarrow$ MLP (GELU, dropout) with residual additions.
  (d) Decoder block: upsampling, concatenation with the encoder skip, followed by two $3{\times}3$ Conv--ReLU layers.
  Red dashed curves indicate skip connections, including U\mbox{-}Net long skips and the residual paths within blocks.}
  \label{fig:arch_blocks}
\end{figure}

The interpolation operator $I(\cdot;\theta)$ follows a U\mbox{-}Net style encoder--decoder tailored for sparse to dense RF-domain reconstruction (Fig.~\ref{fig:arch_blocks}(a)). 
The encoder extracts multi-scale features across four stages (E1--E4) and a bottleneck, while the decoder mirrors this hierarchy (D1--D4) by progressively upsampling and fusing features through long skip connections.

We investigate two interchangeable encoder backbones that provide complementary inductive biases. 
The first is HGNet-V2, a convolution-based design that emphasizes hierarchical feature reuse and local aggregation. 
As illustrated in Fig.~\ref{fig:arch_blocks}(b), each HGNet-V2 block employs multiple parallel $3{\times}3$ Conv--BN--ReLU branches whose outputs are concatenated and projected with a $1{\times}1$ convolution, together with a residual link. 
This structure supports efficient local processing and helps preserve fine-scale continuity in the reconstructed RF signals.

The second backbone is CAFormer, a hybrid convolution--attention design that combines convolutional stems with token-mixing blocks to capture longer-range dependencies. 
As shown in Fig.~\ref{fig:arch_blocks}(c), a CAFormer block applies LayerNorm, a token-mixing stage, a second LayerNorm, and an MLP with GELU and dropout, with residual additions around each sub-layer. 
By modeling broader spatial context, CAFormer can better capture coherent wavefront patterns spanning many channels, which is advantageous under irregular sampling or severe undersampling.

The decoder is shared across both backbones (Fig.~\ref{fig:arch_blocks}(d)). 
At each decoder stage, features are upsampled (by a factor of two), concatenated with the corresponding encoder features, and refined using two $3{\times}3$ Conv--ReLU layers to recover dense RF outputs. 
Both the U\mbox{-}Net long skip connections and the residual paths within blocks (depicted as red dashed curves) facilitate gradient propagation and help preserve fine-scale details.

As introduced in Section~\ref{sec:problem_setup}, sparse inputs are represented by zero-filled masking so the network interface remains fixed across sampling patterns.
We evaluate both HGNet-V2 and CAFormer under the same physics-guided training objective to study how encoder inductive bias affects sparse-to-dense RF reconstruction, with quantitative comparisons reported in Section~\ref{sec:accuracy}.

\section{Data and Training}



This section describes the data simulation and training protocol used to evaluate the proposed sparse-to-dense RF reconstruction framework. We first introduce the RF data simulation procedure, including the radial scanning geometry, transducer and acquisition parameters, and the scatterer configurations used to generate fully sampled RF data. Sparse RF inputs are then derived from the fully sampled data through the decimation and masking strategies described in Section~\ref{sec:random}. We next summarize the training and validation setup, including the network backbones, loss functions, optimization settings, data split, and evaluation metric used throughout the experiments.

\subsection{RF Data Simulation}

\begin{figure}[!ht]
  \centering
  \begin{subfigure}[t]{0.48\linewidth}
    \centering
    \includegraphics[width=\linewidth]{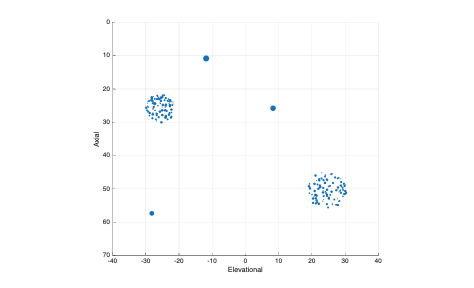}
    \caption{Simulated point targets and speckle regions.}
    \label{fig:sim_scatterers}
  \end{subfigure}\hfill
  \begin{subfigure}[t]{0.48\linewidth}
    \centering
    \includegraphics[width=\linewidth]{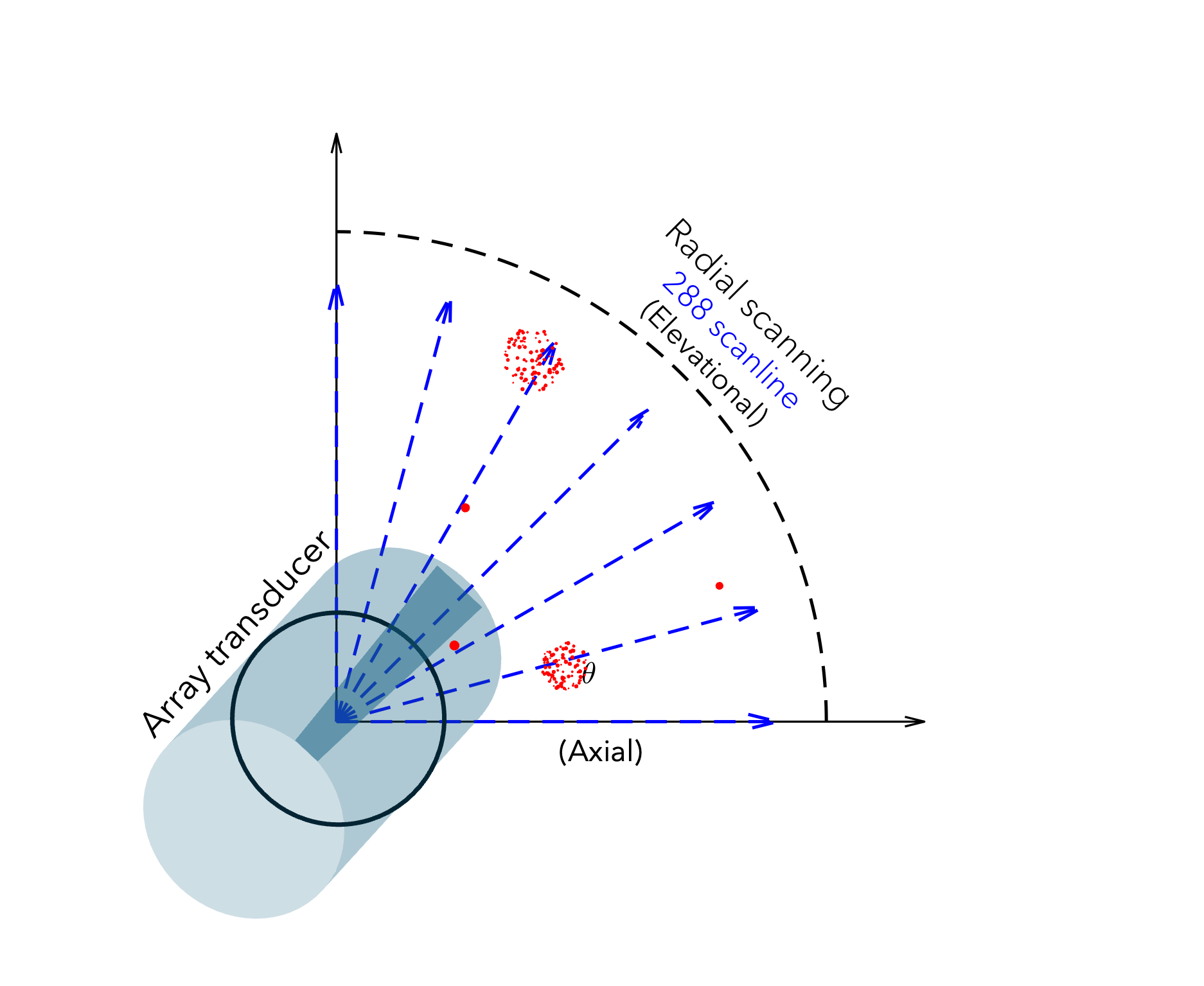}
    \caption{Radial scanning geometry and steering angles.}
    \label{fig:sim_geometry}
  \end{subfigure}
  \caption{Simulation setup. (a) Scatterer layout with point targets and speckle regions. (b) Steered plane-wave transmissions for radial scanning.}
  \label{fig:sim_setup}
\end{figure}

\begin{table}[!t]
\centering
\caption{Array and imaging parameters.}
\label{tab:simulation_table}
\footnotesize
\setlength{\tabcolsep}{4pt}
\renewcommand{\arraystretch}{1.08}

\begin{tabular}{@{}p{0.40\linewidth} p{0.42\linewidth}@{}}
\toprule
\textbf{Array parameters} & \textbf{Details} \\
\midrule
Transducer type          & Linear \\
Number of elements       & 128 \\
Speed of sound           & 1540 m/s \\
Array pitch              & 0.43 mm \\
Element width            & 0.42 mm \\
Element height           & 7 mm \\
Elevation focus          & 5 mm \\
Transmit frequency       & 6.5 MHz \\

\midrule
\textbf{Imaging parameters} & \textbf{Details} \\
\midrule
Radial scanning angle range & $-45^\circ$ to $45^\circ$ \\
Number of scanlines         & 288 \\
Radial scan interval        & $0.3128^\circ$ \\
TX pulse cycles             & 1 \\
Sampling rate               & 26 MHz \\
View depth                  & 65 mm \\
Dynamic range               & 60 dB \\

\bottomrule
\end{tabular}

\vspace{-5pt}
\end{table}
In this paper, we validate our network to facilitate the volumetric scanning approach in our previous study \parencite{song2022novel}, in which each radial plane is scanned by a linear array to establish a target volume using synthetic radial aperture focusing technique. Any reduction in the number of scanning events would be converted into gain in overall scanning speed and computational/power efficiency. We simulated the RF data using Field II~ \parencite{jensen1997field} with a 128-element linear array. A total of 288 radial planes were scanned over $-45^\circ$ to $45^\circ$ range with radial spacing of $0.3128^\circ$ (Fig.~\ref{fig:sim_geometry}). 
Each radial scanning angle provides a unique radial view of the target volume, and synthetic radial aperture focusing technique reconstructs the target volume at high spatial resolution \parencite{song2022novel}. 

\begin{figure}[H]
  \centering
  \includegraphics[width=0.60\linewidth]{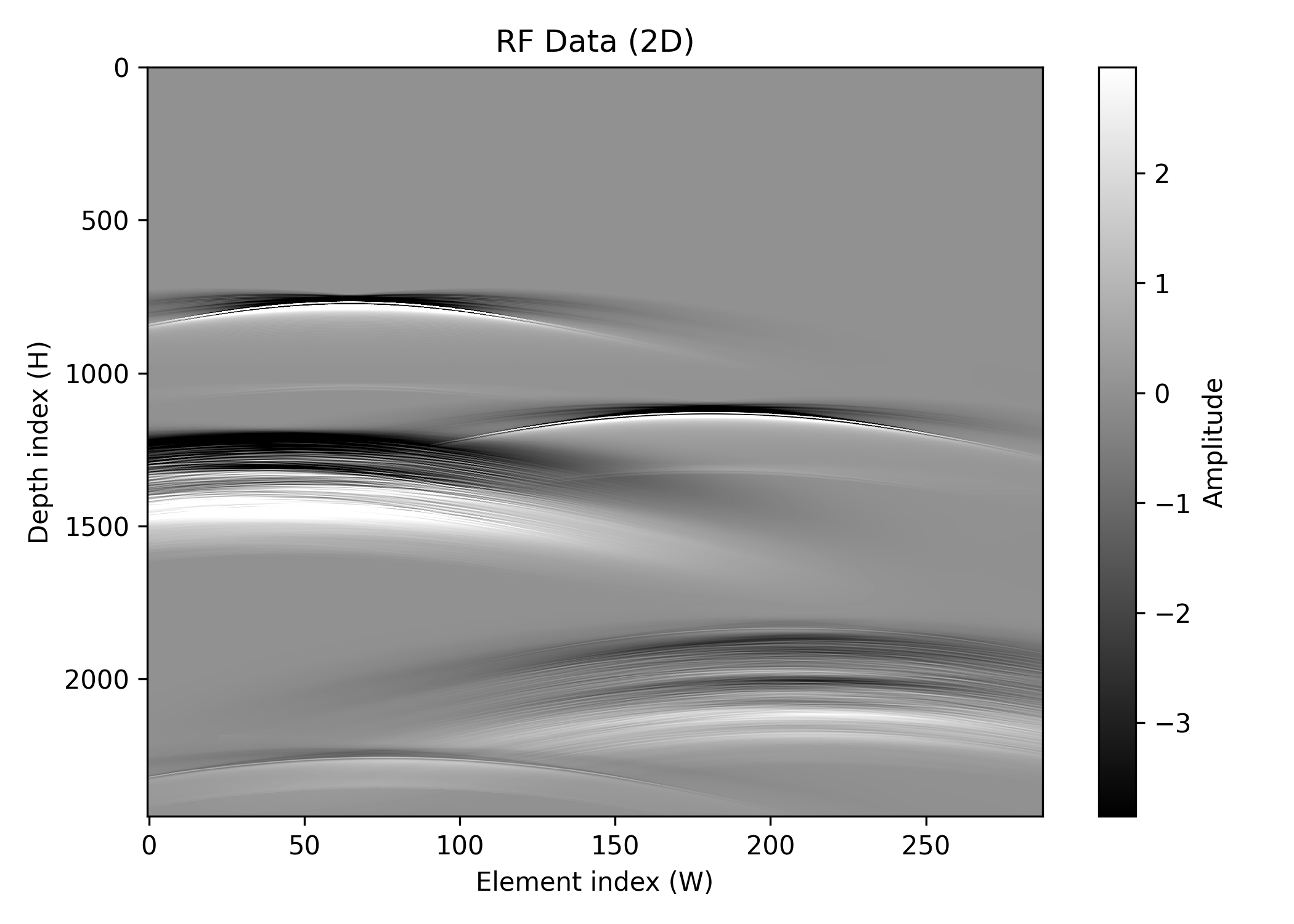}
  \caption{Example simulated RF channel data (channel--time matrix) from one transmit, shown before beamforming.}
  \label{fig:sim_rf_example}
\end{figure}

We designed each simulation to contain 2--6 point targets and 1--3 hyperechoic mass targets with 2-6 mm radius, randomly placed within the FOV (Fig.~\ref{fig:sim_scatterers}). Each element was excited by a single-cycle 6.5 MHz sinusoidal pulse, and the received RF signals were sampled at 26 MHz. As hyperechoic features, each mass target contained 100-point scatterers distributed within the spherical volume, while each point target is represented by a single scatterer. The intensity of target scatterers was randomly assigned from a range of 15–25.
In addition to the basic structures, background scatterers were added in the synthetic phantom generation with lower intensities sampled between 1–5. 
An example simulated RF channel--time matrix is shown in Fig.~\ref{fig:sim_rf_example}.
After acquisition, the full channel RF data were normalized to prepare for the training dataset using global z-score standardization.

\subsection{Training and Evaluation Setup}

\begin{table}[!t]
\centering
\caption{Training and evaluation configuration.}
\label{tab:train_settings}
\footnotesize
\setlength{\tabcolsep}{5pt}
\renewcommand{\arraystretch}{1.08}
\begin{tabular}{@{}p{0.34\linewidth} p{0.58\linewidth}@{}}
\toprule
\textbf{Parameter} & \textbf{Setting} \\
\midrule
Training loader          & Batch size $=16$ \\
Validation loader        & Batch size $=16$ \\
Test loader              & Batch size $=16$ \\
Optimizer                & Adam \\
Learning rate            & $1\times10^{-4}$ \\
Epochs                   & 100 \\
Random seed              & 1234 \\
Loss (RF domain)         & $\ell_1$ loss \\
Loss (BF domain)         & MSE after beamforming \\
BF loss weighting        & EMA-based adaptive weighting: $k=0.20$, $\beta=0.90$, $\varepsilon=10^{-8}$, and $\lambda_0=2\times10^{-4}$ \\
Encoder backbone         & HGNet-V2 / CAFormer \\
Data split               & 8:1:1 train--validation--test split (2,048/256/256) \\
Workers                  & 20 \\
Evaluation metric        & Mean SSIM \\
Hardware                 & NVIDIA GH200, 120\,GB HBM \\
\bottomrule
\end{tabular}
\vspace{-6pt}
\end{table}

Table~\ref{tab:train_settings} summarizes the training and evaluation configuration for the proposed sparse-to-dense RF reconstruction framework. The models are trained in a fully supervised setting for 100 epochs using Adam with a batch size of 16, a learning rate of $1\times10^{-4}$, and a fixed random seed of 1234. Two encoder variants are evaluated: HGNet-V2 and CAFormer.

The training objective combines an RF-domain $\ell_1$ loss with a physics-guided MSE loss computed after beamforming. Their relative weight is adapted during training using the EMA-based weighting scheme. Specifically, $k=0.20$ sets the target weighted beamforming-domain contribution to $20\%$ relative to the RF-domain loss. The EMA smoothing factor is set to $\beta=0.90$, with an initial weight of $\lambda_0=2\times10^{-4}$ and a numerical stabilizer of $\varepsilon=10^{-8}$.

The simulated speckle dataset contains 2,560 samples and is divided into 2,048 training samples, 256 validation samples, and 256 test samples. The validation set is used exclusively for model selection and hyperparameter tuning, including selection of the loss configuration and random-skip range, whereas the held-out test set remains unseen until the final evaluation. Data loading is performed using 20 parallel workers. All experiments are run on a single NVIDIA GH200 GPU with 120\,GB HBM. Final evaluation results are reported as the mean SSIM across the 256 held-out test samples, whereas ablation results used for model selection are reported on the 256-sample validation set.

\section{Results}
We first describe the simulation protocol used to generate paired dense and sparse RF data, including the scatterer layout and scanning geometry, and then summarize the training and test settings. We next report RF-domain reconstructions and grating-lobe suppression on standardized point-target scenes for quantitative evaluation of mainlobe-to-gratinglobe ratio (MGR) metric. We then present image-level reconstruction accuracy across sparsity levels using SSIM \parencite{wang2004ssim} , followed by qualitative comparisons that highlight artifact suppression under both ID and OOD decimation. Finally, we evaluate robustness to varying sampling rates with randomized masking, and report ablations, noise robustness tests over the interferences from background scatterers and noise, and generalization results on synthetic phantom data. Unless otherwise stated, results are reported as mean SSIM on the held-out set, and figures and tables show the full test distribution.

\subsection{RF-domain Reconstruction}
\label{subsec:rf_reconstruction}

As an initial qualitative evaluation of the RF-domain component of our
hybrid objective, we examined the reconstructed RF data across different
decimation factors, focusing on the continuity and coherence of the recovered
wavefronts.

\begin{figure}[H]
  \centering
  \includegraphics[width=\linewidth]{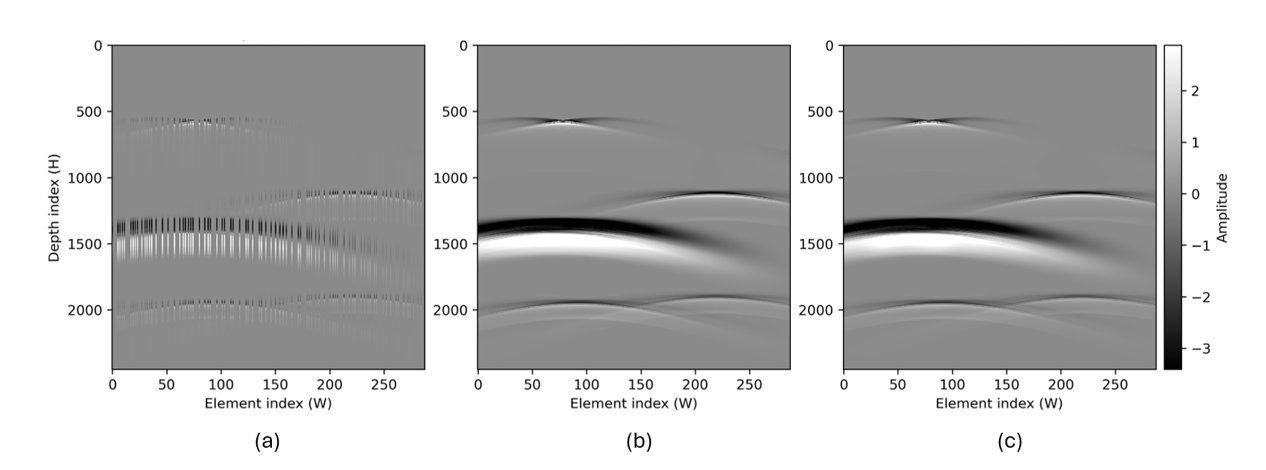}
  \caption{Representative training example in the RF domain.
    (a) Sparse RF input generated using random-skip masking with a maximum
    skip length of $R=7$.
    (b) Dense ground-truth RF data $V_{\text{full}}$.
    (c) Network prediction $\hat{V}$.}
  \label{fig:rf_train_vis}
\end{figure}

Figure~\ref{fig:rf_train_vis} shows a representative example from the
training set, including the sparse RF input, dense ground truth, and network
prediction.

Compared with the masked input in Figure~\ref{fig:rf_train_vis}(a), the
prediction in Figure~\ref{fig:rf_train_vis}(c) restores the missing channels
and recovers continuous hyperbolic wavefronts that closely resemble those in
the dense target shown in Figure~\ref{fig:rf_train_vis}(b). The reconstruction
also substantially reduces the periodic striping introduced by the masking
operation.

\begin{figure}[H]
  \centering
  \includegraphics[width=\linewidth]{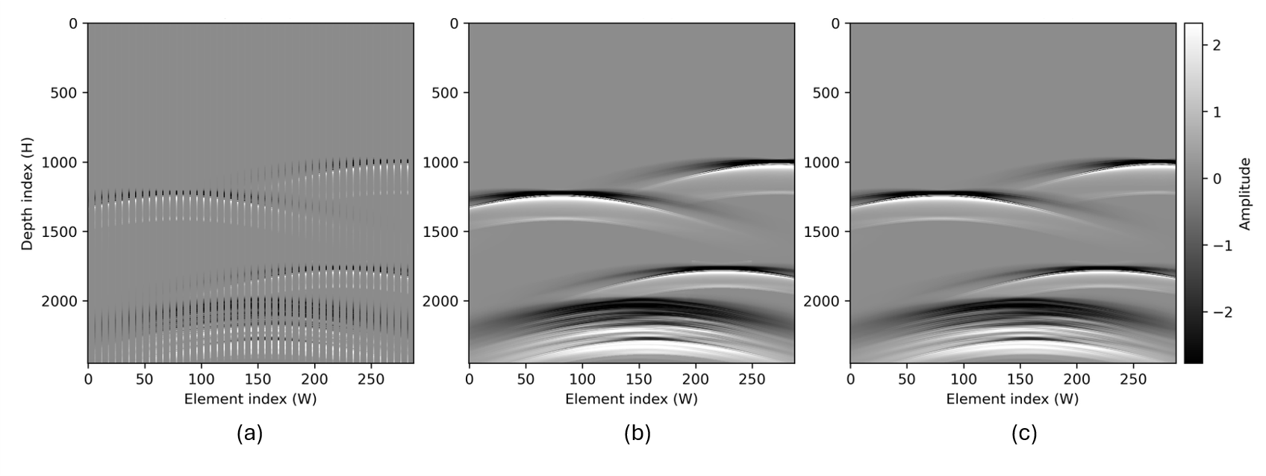}
  \caption{Representative held-out test example at a fixed decimation factor
    of $\times 6$.
    (a) Regularly decimated RF input.
    (b) Dense ground-truth RF data $V_{\text{full}}$.
    (c) Network prediction $\hat{V}$.}
  \label{fig:rf_test_vis}
\end{figure}

Figure~\ref{fig:rf_test_vis} presents a held-out test example obtained using
a fixed regular decimation factor of $\times 6$.

Although this regular test-time sampling pattern differs from the randomized
masks used during training, it remains within the sparsity range represented
during training. The model reconstructs coherent echoes across channels and
preserves the depth-dependent wavefront geometry, indicating generalization
beyond the training samples and masking layouts.

These qualitative RF-domain results indicate that the reconstructed channels
maintain sufficient phase and amplitude consistency for coherent summation.
The corresponding beamforming-domain effects are quantified later using
angular intensity profiles and the main-to-grating-lobe ratio (MGR).

\subsection{Image Reconstruction Accuracy and Robustness: Quantitative Evaluation}
\label{sec:accuracy}

To assess image reconstruction accuracy and robustness under varying acquisition budgets, we evaluate beamformed-image SSIM across a wide range of decimation factors on a held-out test set.
This experiment has two goals: (i) to quantify how beamforming-domain supervision and EMA loss balancing affect reconstruction quality, and (ii) to test robustness as sparsity increases beyond the training range.
During training, we use random-skip masking with a maximum skip length $R=7$, which corresponds to decimation factors up to $\times 8$.
We therefore treat $\times 2$--$\times 8$ as ID and $\times 9$--$\times 13$ as OOD relative to the training sparsity range.

\begin{figure}[!htbp]
\centering
\includegraphics[width=0.95\linewidth]{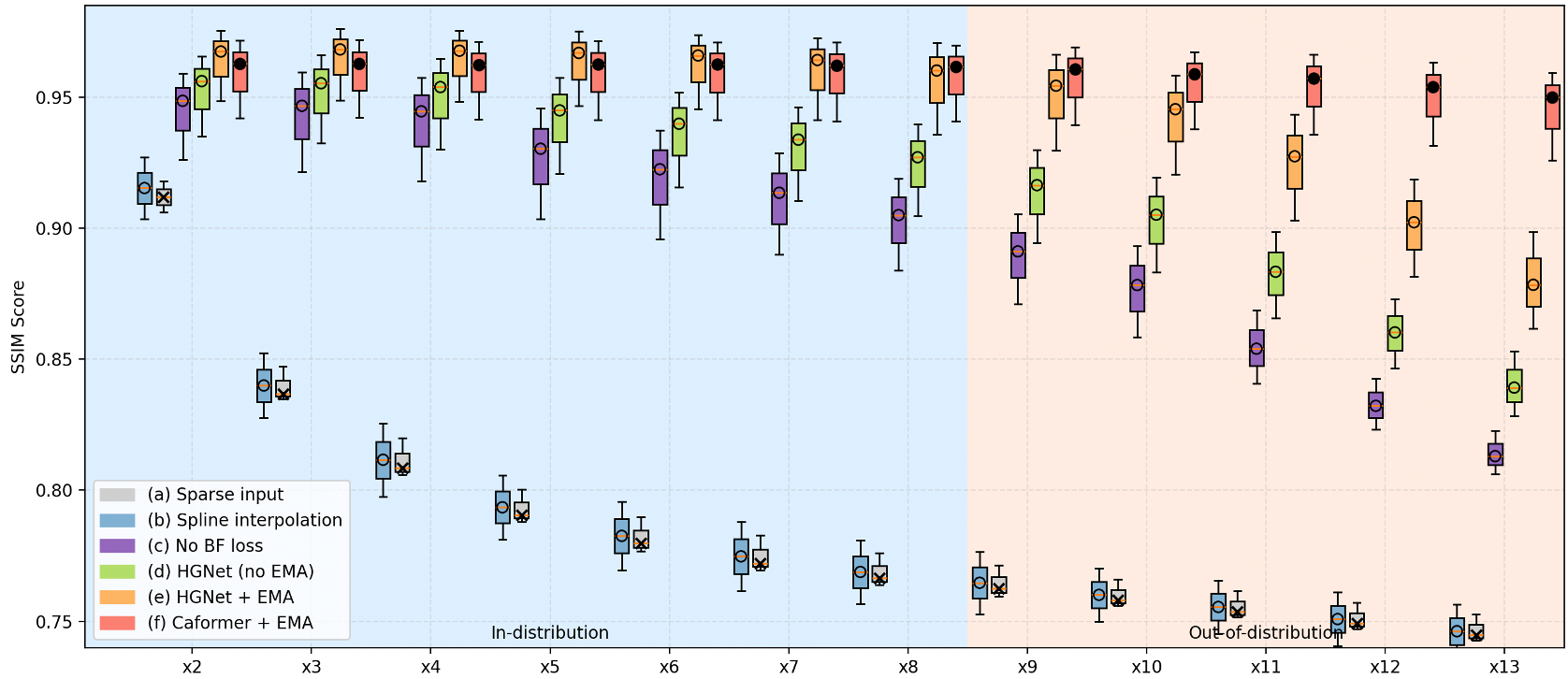}
\caption{SSIM across decimation levels. In-distribution (ID) covers $\times 2$--$\times 8$, and out-of-distribution (OOD) covers $\times 9$--$\times 13$.
We compare six settings: (a) sparse input, (b) spline interpolation, (c) no BF loss, (d) HGNet-V2 (no EMA), (e) HGNet-V2 + EMA, and (f) CAFormer + EMA.
Higher SSIM values indicate better reconstruction quality with respect to the densely sampled ground truth image.}
\label{fig:ssim_box}
\end{figure}

We evaluate all methods on the held-out test set using regular, fixed decimation patterns, while training uses random-skip masking.
Figure~\ref{fig:ssim_box} summarizes SSIM distributions across decimation levels.
As expected, SSIM decreases as sparsity increases, but the rate of decline and the dispersion depend strongly on the training objective and backbone.
Directly beamforming the sparse input (a) performs worst across all decimations.
We include spline interpolation (b) as a classical nonlearning baseline that fills missing RF channels along the aperture using a natural cubic spline.
Although it enforces smoothness, it does not explicitly preserve the coherent channel relationships required for coherent DAS beamforming, so its improvement over (a) remains limited.

Beamforming-domain supervision yields clear gains over RF-only training.
Compared with the no-BF-loss baseline (c), introducing beamforming-loss guided training (d) improves mean SSIM by 2.23\%, supporting the motivation that image-formation-aware supervision reduces beamforming-visible errors that can accumulate during coherent DAS.
EMA-based adaptive weighting further improves mean SSIM by 3.25\% and tightens the SSIM distributions (e), consistent with more stable multi-objective optimization.
Finally, replacing the convolutional encoder with the hybrid conv--attention encoder improves robustness at higher decimation, providing an additional 3.83\% gain in OOD ($\times 9$--$\times 13$) mean SSIM for CAFormer + EMA (f) relative to HGNet-V2 + EMA (e).
Overall, the best-performing configuration (f) maintains strong performance across $\times 2$--$\times 13$, with the most gradual degradation under severe undersampling.

In the next subsection, we complement these aggregate statistics with qualitative visual comparisons to illustrate the characteristic artifacts at different decimation levels and how the proposed components reduce them.

\subsection{Image Reconstruction Accuracy and Robustness: Qualitative Evaluation}

\begin{figure}[!htbp]
\centering
\includegraphics[width=\textwidth]{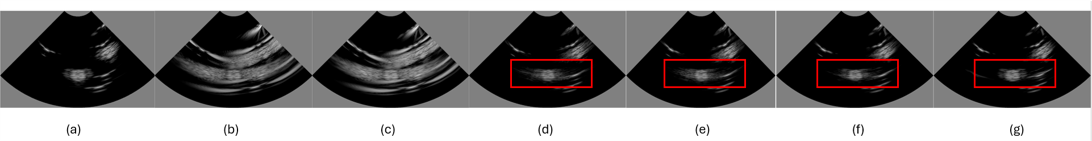}
\caption{Qualitative comparison at $\times 6$ decimation (in-distribution).
(a) Full RF (ground truth); (b) sparse input + DAS; (c) spline interpolation; (d) RF-only loss; (e) RF + BF loss; (f) RF + BF loss + EMA; (g) RF + BF loss + EMA (CAFormer).
Red boxes highlight regions where physics-guided supervision reduces grating-lobe artifacts and better preserves structures.}
\label{fig:vis_x6}
\end{figure}

\begin{figure}[!t]
\centering
\includegraphics[width=\textwidth]{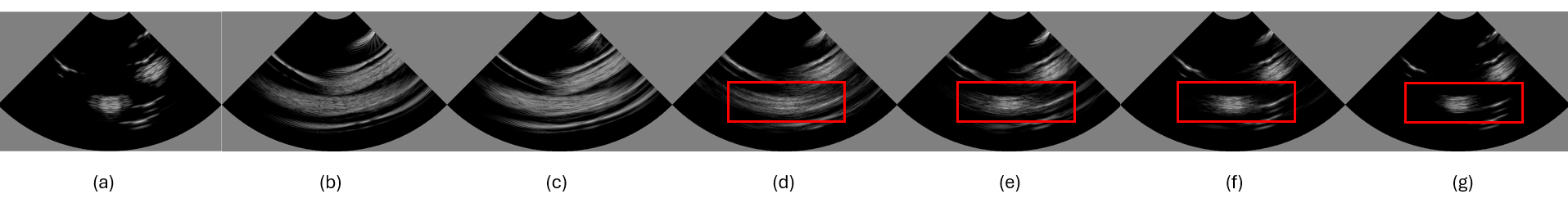}
\caption{Qualitative comparison at $\times 13$ decimation (out-of-distribution).
(a) Full RF (ground truth); (b) sparse input + DAS; (c) spline interpolation; (d) RF-only loss; (e) RF + BF loss; (f) RF + BF loss + EMA; (g) RF + BF loss + EMA (CAFormer).
Red boxes highlight regions where physics-guided supervision mitigates grating-lobe artifacts under severe undersampling.}
\label{fig:vis_x13}
\end{figure}

Figures~\ref{fig:vis_x6} and~\ref{fig:vis_x13} provide qualitative comparisons at $\times 6$ (ID) and $\times 13$ (OOD) decimation, complementing the SSIM trends in Fig.~\ref{fig:ssim_box}.
At $\times 6$ (Fig.~\ref{fig:vis_x6}), directly beamforming the sparse input (b) exhibits prominent grating lobes and blurred boundaries inside the highlighted region, and spline interpolation (c) offers limited improvement.
Training with RF-only supervision (d) reduces some artifacts but still leaves noticeable sidelobes and edge smearing, indicating that low RF-domain error does not guarantee clean post-beamforming appearance.
In contrast, introducing beamforming-domain supervision (e) visibly suppresses grating lobes and sharpens boundaries, and EMA-based balancing (f) further reduces residual clutter and stabilizes the reconstruction.
The CAFormer variant (g) yields the cleanest appearance among learned methods, with the weakest residual sidelobes and the most consistent boundary definition.

Under severe undersampling at $\times 13$ (Fig.~\ref{fig:vis_x13}), the differences become more pronounced.
Classical baselines (b,c) and RF-only training (d) show strong structured artifacts and loss of detail, while beamforming-guided training (e,f) better preserves structures and suppresses grating-lobe patterns in the boxed region.
Consistent with the quantitative OOD gap, the CAFormer model (g) degrades more gracefully and maintains clearer boundaries with fewer coherent artifacts than the convolutional alternative.

These visual results illustrate how beamforming-domain supervision, stable loss balancing, and long-range context modeling translate into reduced grating lobes and improved perceptual quality, especially at high decimation.

\subsection{Grating-lobe Suppression}
\label{subsec:mgr_analysis}

To evaluate whether the proposed reconstruction suppresses grating-lobe
artifacts after beamforming, we analyzed the angular responses of isolated
point targets at five radial depths. The analyzed regions were approximately
3~mm wide and centered at 13.30, 26.33, 39.36, 52.39, and 65.13~mm relative
to the transducer surface.

For each region, the linear envelope magnitude was summed along the depth
direction and normalized by its maximum value to obtain an angular response.
The maximum response was defined as the main-lobe amplitude
$A_{\mathrm{main}}$. After excluding an angular neighborhood of
$\pm3^{\circ}$ around the main-lobe peak, the strongest remaining off-axis
response was defined as the grating-lobe amplitude $A_{\mathrm{gr}}$.
The main-to-grating-lobe ratio (MGR) was calculated as

\[
\mathrm{MGR}_{\mathrm{dB}}
=
20\log_{10}
\left(
\frac{A_{\mathrm{main}}}{A_{\mathrm{gr}}}
\right).
\]

A higher MGR indicates stronger main-lobe dominance and better suppression
of off-axis grating-lobe responses.

\begin{figure}[!htbp]
    \centering

    \begin{subfigure}[t]{0.48\linewidth}
        \centering
        \includegraphics[width=\linewidth]{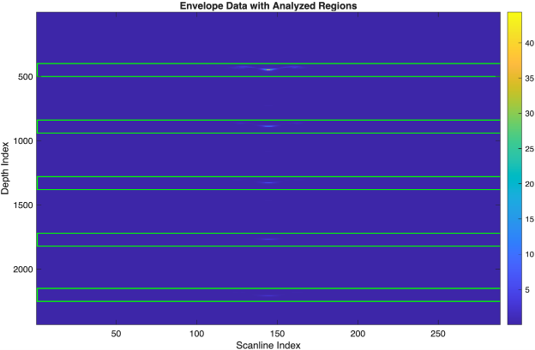}
        \caption{Beamformed envelope image with the five analyzed regions.}
        \label{fig:gl_regions}
    \end{subfigure}
    \hfill
    \begin{subfigure}[t]{0.48\linewidth}
        \centering
        \includegraphics[width=\linewidth]{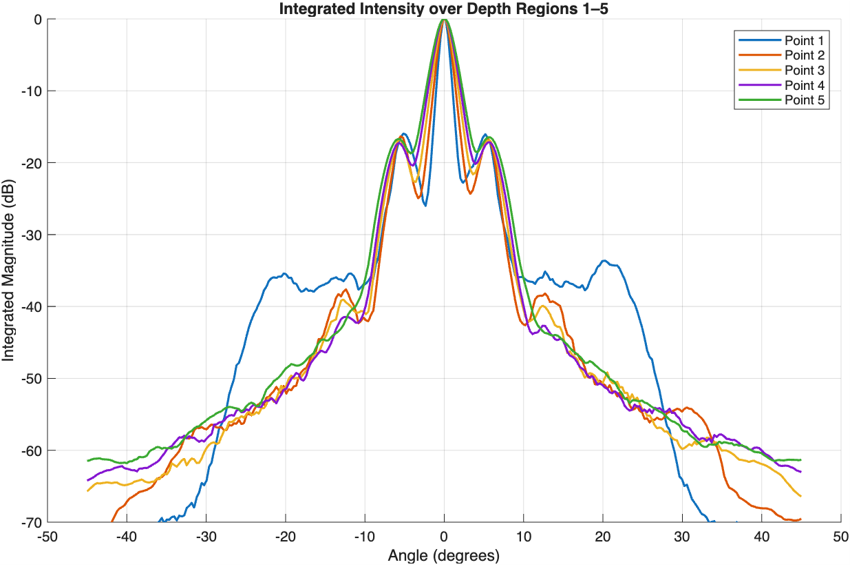}
        \caption{Fully sampled ground truth.}
        \label{fig:gl_gt}
    \end{subfigure}

    \vspace{6pt}

    \begin{subfigure}[t]{0.48\linewidth}
        \centering
        \includegraphics[width=\linewidth]{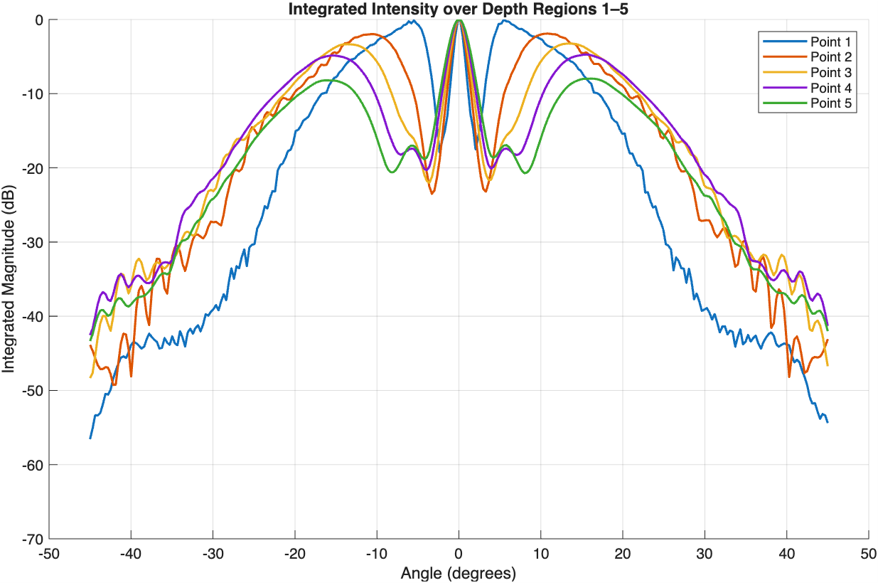}
        \caption{Sparse input at $\times6$ decimation.}
        \label{fig:gl_sparse}
    \end{subfigure}
    \hfill
    \begin{subfigure}[t]{0.48\linewidth}
        \centering
        \includegraphics[width=\linewidth]{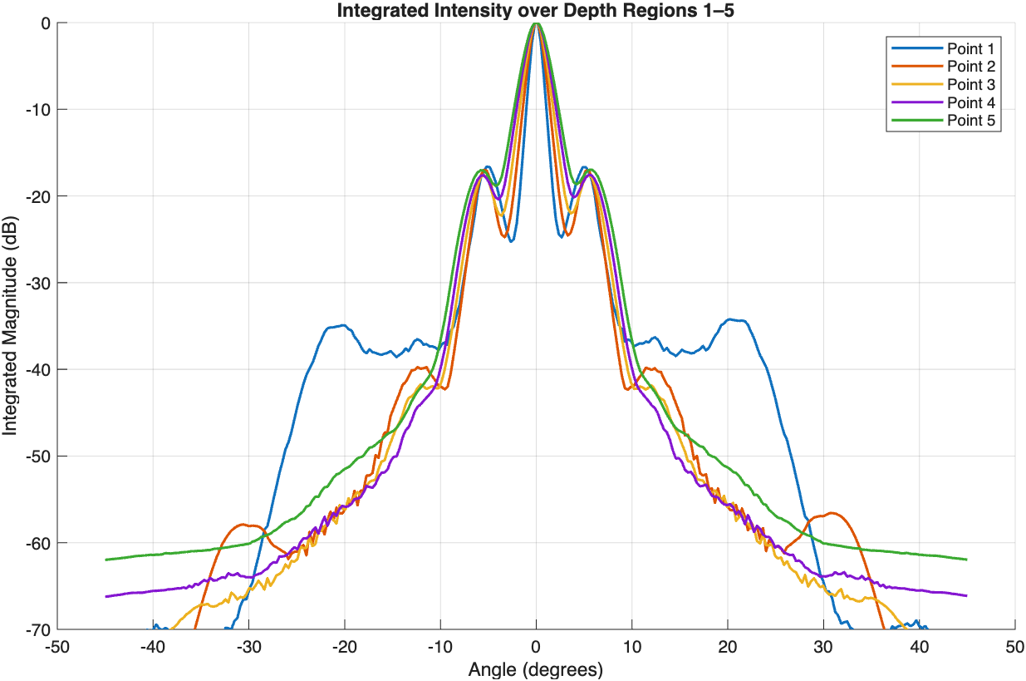}
        \caption{Reconstructed prediction at $\times6$ decimation.}
        \label{fig:gl_pred}
    \end{subfigure}

    \caption{Angular-response analysis at $\times6$ decimation.
    (a) The five analyzed regions, centered at depths of 13.30, 26.33,
    39.36, 52.39, and 65.13~mm.
    (b)--(d) Normalized angular responses for the fully sampled ground truth,
    sparse input, and reconstructed prediction, respectively.}
    \label{fig:globe_analysis}
\end{figure}

Figure~\ref{fig:globe_analysis} shows that sparse sampling produces elevated
off-axis responses across the five analyzed depths. After reconstruction,
the angular profiles become substantially closer to the fully sampled
reference, with a compact main lobe and reduced off-axis energy.

\begin{table}[!htbp]
    \centering
    \caption{MGR across decimation factors. Values are the mean $\pm$
    standard deviation across the five analyzed depth regions ($n=5$).
    The fully sampled reference has an MGR of
    $16.44 \pm 1.26$~dB for all decimation factors.}
    \label{tab:mgr_across}
    \begin{tabular}{ccc}
        \toprule
        Decimation factor &
        Decimated MGR (dB) &
        Reconstructed MGR (dB) \\
        \midrule
        $\times2$  & $14.03 \pm 2.69$ & $16.98 \pm 1.45$ \\
        $\times4$  & $ 7.85 \pm 4.81$ & $16.58 \pm 1.36$ \\
        $\times6$  & $ 3.58 \pm 3.00$ & $15.96 \pm 1.04$ \\
        $\times8$  & $ 2.25 \pm 1.65$ & $15.75 \pm 0.95$ \\
        $\times10$ & $ 1.53 \pm 1.15$ & $15.05 \pm 0.72$ \\
        $\times12$ & $ 1.10 \pm 1.03$ & $14.39 \pm 1.20$ \\
        \bottomrule
    \end{tabular}
\end{table}

As shown in Table~\ref{tab:mgr_across}, the mean MGR of the decimated input
decreases from 14.03~dB at $\times2$ to 1.10~dB at $\times12$. In contrast,
the reconstructed RF data maintain mean MGR values between 14.39 and
16.98~dB. At the $\times6$ setting shown in
Figure~\ref{fig:globe_analysis}, reconstruction increases the mean MGR from
$3.58 \pm 3.00$~dB to $15.96 \pm 1.04$~dB, approaching the fully sampled
reference value of $16.44 \pm 1.26$~dB.

These results demonstrate that the proposed reconstruction preserves
beamforming-relevant angular responses and substantially suppresses
grating-lobe artifacts under sparse acquisition.

\subsection{Robustness to Sparsity via Randomized Sampling}
\label{sec:random_skip_rate}

This experiment evaluates robustness to unknown spatial sampling rates by comparing models trained under a fixed decimation factor with a single model trained using randomized sampling (random skip).
In practical systems, the effective receive sparsity can vary across probes, sessions, or operating modes, so a model that performs well only at one trained decimation may be brittle when the sampling rate changes.
We therefore test whether random-skip training produces a single model that maintains consistently high image quality across multiple decimation levels.

Table~\ref{tab:sparsity} reports beamformed-image SSIM for models trained on a fixed decimation ($\times 2$, $\times 4$, or $\times 8$) and evaluated at $\times 2$, $\times 4$, and $\times 8$.

\begin{table}[!htbp]
  \centering
  \caption{SSIM for models trained with fixed or randomized decimation,
  evaluated across $\times2$, $\times4$, and $\times8$ test decimation
  levels. The best result in each row is shown in bold.}
  \label{tab:sparsity}
  \vspace{4pt}
  \begin{tabular}{@{}lcccc@{}}
    \toprule
    \diagbox{Test}{Train}
    & \textbf{Train on $\times2$}
    & \textbf{Train on $\times4$}
    & \textbf{Train on $\times8$}
    & \textbf{Random skip ($R=7$)} \\
    \midrule
    \textbf{Test on $\times2$}
    & \textbf{0.982905} & 0.927430 & 0.832716 & 0.963876 \\
    \textbf{Test on $\times4$}
    & 0.950306 & \textbf{0.978986} & 0.825197 & 0.962189 \\
    \textbf{Test on $\times8$}
    & 0.933108 & 0.937244 & \textbf{0.967669} & 0.962389 \\
    \bottomrule
  \end{tabular}
\end{table}

As expected, fixed-rate models perform best when the test decimation matches the training decimation (diagonal entries) but degrade under mismatch.
For example, the $\times 2$ model drops from 0.9829 at $\times 2$ to 0.9503 at $\times 4$ and 0.9331 at $\times 8$, while the $\times 4$ model drops to 0.9274 at $\times 2$ and 0.9372 at $\times 8$.
Although the $\times 8$ model generalizes better to $\times 8$ (0.9677), it performs substantially worse at lower decimation (0.8327 at $\times 2$ and 0.8252 at $\times 4$), indicating that training on a single high-sparsity regime can bias the model toward that regime.

In contrast, the random-skip model achieves consistently high SSIM across all test decimations, with 0.9639 at $\times2$, 0.9622 at $\times4$, and 0.9624 at $\times8$.
While it may not always exceed the best matched fixed-rate model, it avoids the severe failures under mismatch and provides reliable performance across sparsity levels.
These results support random-skip training as an effective strategy for improving robustness to varying receive-channel layouts and sampling rates.
Next, we perform ablation studies to isolate the contributions of key design choices, including the beamforming-domain loss formulation and the random-skip range used to generate training masks.

\subsection{Ablation Tests and Studies}
We study two design choices that most directly affect image quality and robustness.
First, we ablate the beamforming-domain loss to examine how different image-domain objectives guide the network toward beamforming-relevant error reduction.
Second, we ablate the random-skip setting used to create masked inputs, since the maximum skip length controls the sparsity patterns seen during training and therefore the trade-off between in-range accuracy and robustness to higher decimation at test time. All ablation results in Tables~\ref{tab:bf_loss_ablation} and~\ref{tab:ablation_skip_rate} are reported on the validation set, whereas the independent test set is reserved for the final model evaluation.

\FloatBarrier
\subsubsection{Tests on Beamforming-domain Loss Functions}
\label{sec:bf-loss-ablation}

We compare three beamforming-domain losses: mean squared error (MSE), multi-scale SSIM (MS\_SSIM), and a perceptual loss computed on deep features.
MSE directly penalizes pixel-wise intensity differences in the beamformed image, MS\_SSIM emphasizes structural similarity across multiple scales, and the perceptual loss encourages similarity in a learned feature space.

Across all tested decimation factors, MSE achieves the highest mean SSIM on the validation set (Table~\ref{tab:bf_loss_ablation}).

\begin{table}[!htbp]
  \centering
  \caption{Ablation of the beamforming-domain loss based on mean SSIM on the validation set.
  Rows indicate the loss used in $\mathcal{L}_{\mathrm{BF}}$, and columns indicate the validation decimation factor.
  The best result in each column is shown in \textbf{bold}.}
  \label{tab:bf_loss_ablation}
  \vspace{4pt}
  \begin{tabular}{@{}lccccc@{}}
    \toprule
    \textbf{BF-domain loss}
    & \textbf{$\times2$}
    & \textbf{$\times4$}
    & \textbf{$\times6$}
    & \textbf{$\times8$}
    & \textbf{$\times10$} \\
    \midrule
    MS\_SSIM
    & 0.948409
    & 0.950341
    & 0.947218
    & 0.939233
    & 0.921916 \\
    MSE
    & \textbf{0.962686}
    & \textbf{0.962372}
    & \textbf{0.962467}
    & \textbf{0.961501}
    & \textbf{0.958763} \\
    Perceptual
    & 0.949629
    & 0.953274
    & 0.952739
    & 0.945869
    & 0.938023 \\
    \bottomrule
  \end{tabular}
\end{table}
\FloatBarrier

In contrast, MS\_SSIM and perceptual losses yield consistently lower SSIM, with the gap becoming more pronounced at higher sparsity.
A plausible reason is that MSE more directly penalizes intensity deviations that reflect coherent summation artifacts, such as elevated sidelobes and blur, which dominate quality degradation under severe undersampling.
By comparison, MS\_SSIM and perceptual objectives emphasize structural or feature-level similarity and may tolerate intensity biases that reduce SSIM, particularly when the beamformed image contains localized artifacts.
Based on these results, we use beamforming-domain MSE for $\mathcal{L}_{\mathrm{BF}}$ in all other experiments.

\FloatBarrier
\subsubsection{Tests on Maximum Skip Length in Random-skip Masking}
\label{sec:rand-skip-ablation}

We compare three random-skip settings by varying the maximum skip length $R$, which determines the longest consecutive missing-channel run seen during training. A smaller $R$ biases training toward milder sparsity patterns, whereas a larger $R$ exposes the model to more aggressive channel gaps and may improve robustness at higher decimation.

Table~\ref{tab:ablation_skip_rate} shows that $R$ controls a clear trade-off between low-decimation accuracy and robustness at higher decimation factors.

\begin{table}[!htbp]
  \centering
  \caption{Ablation of the maximum skip length $R$ in random-skip masking based on mean SSIM on the validation set.
  Rows indicate the training setting (Skip 1--$R$), and columns indicate the validation decimation factor.
  The best result in each column is shown in \textbf{bold}.}
  \label{tab:ablation_skip_rate}
  \vspace{4pt}
  \begin{tabular}{@{}lccccc@{}}
    \toprule
    \textbf{Training skip}
    & \textbf{$\times2$}
    & \textbf{$\times4$}
    & \textbf{$\times6$}
    & \textbf{$\times8$}
    & \textbf{$\times10$} \\
    \midrule
    Skip 1--3
    & \textbf{0.964478}
    & \textbf{0.962860}
    & 0.946192
    & 0.912069
    & 0.881246 \\
    Skip 1--7
    & 0.962686
    & 0.962372
    & \textbf{0.962467}
    & \textbf{0.961501}
    & 0.958763 \\
    Skip 1--9
    & 0.950026
    & 0.954631
    & 0.955741
    & 0.958099
    & \textbf{0.959251} \\
    \bottomrule
  \end{tabular}
\end{table}
\FloatBarrier

With a smaller $R$ (Skip 1--3), the model achieves the highest SSIM at low decimation factors ($\times2$ and $\times4$), suggesting that emphasizing short missing runs improves performance when sparsity is mild. However, its performance drops sharply at higher decimation factors ($\times8$ and $\times10$), indicating limited ability to bridge long gaps that are rarely observed during training when $R$ is small.

Conversely, a larger $R$ (Skip 1--9) improves performance at high decimation and achieves the highest SSIM at $\times10$, but slightly reduces performance in the easier low-decimation cases. Across the tested settings, $R=7$ provides a balanced choice, maintaining strong SSIM from $\times2$ to $\times10$ while avoiding the substantial degradation observed when the training skip range is too narrow. Based on these validation results, we use $R=7$ in the remaining experiments.

\subsection{Tests on Noise Robustness}
In practice, received RF measurements are corrupted by electronic and acoustic noise. 
To evaluate noise robustness, we inject additive white Gaussian noise (AWGN) into the \emph{test} inputs only, while keeping training noise-free. 
For a clean RF frame $V$ (linear amplitude units), we estimate the per-frame signal power as $P_{\text{sig}}=\mathrm{mean}(V^2)$. 
Given a target $\mathrm{SNR}_{\mathrm{dB}}$, we set the noise standard deviation as
\[
\sigma \;=\; \sqrt{\frac{P_{\text{sig}}}{10^{\mathrm{SNR}_{\mathrm{dB}}/10}}}\,,
\]
and form the noisy input $\tilde V = V + n$ with $n \sim \mathcal{N}(0,\sigma^2)$. 
We then evaluate reconstruction quality using mean SSIM on the held-out test set for several fixed decimation factors and SNR levels.

\begin{figure}[!t]
\centering
\includegraphics[width=0.95\linewidth]{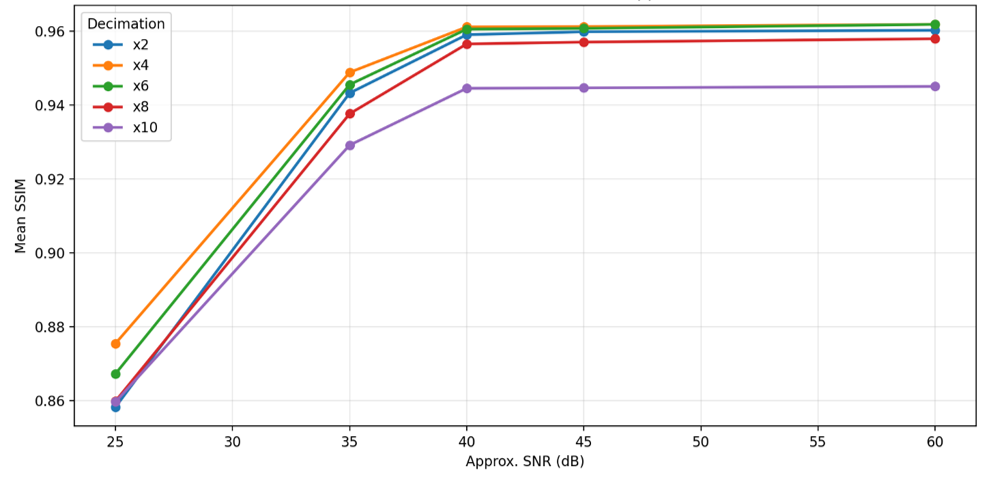}
\caption{Mean SSIM under additive white Gaussian noise injected at test time. Curves correspond to fixed decimation factors; the x-axis shows the approximate input SNR (dB).}
\label{fig:noise_test}
\end{figure}

Figure~\ref{fig:noise_test} shows that performance improves monotonically as SNR increases across all decimation factors, with the largest gains occurring between 25\,dB and 40\,dB. 
Beyond roughly 40\,dB, SSIM saturates and the curves become nearly flat, indicating that reconstruction quality is then limited more by undersampling than by noise. 
As expected, higher decimation factors are consistently more sensitive to noise: at low SNR (25\,dB), the SSIM gap between mild decimation (e.g., $\times2$--$\times6$) and stronger decimation (e.g., $\times8$--$\times10$) is more pronounced, while at high SNR (40--60\,dB) the curves converge toward their noise-free performance. 
Overall, these results indicate that the proposed model remains stable under moderate measurement noise and that denoising is not the primary bottleneck once SNR exceeds approximately 40\,dB.

\subsection{Generalization Tests with Resolution Phantom-like Data}
\label{sebsec:GeneralizationTest}
\begin{figure}[!htbp]
  \centering
  \includegraphics[width=\linewidth]{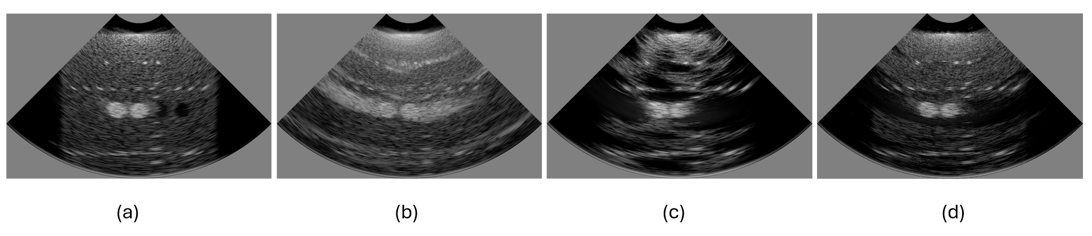}
  \caption{Generalization test on a resolution phantom-like dataset under domain shift.
  (a) Ground-truth beamformed image from fully sampled RF.
  (b) Direct beamforming of the $\times6$ decimated sparse input (baseline).
  (c) Zero-shot prediction of our method (no fine-tuning).
  (d) Our method after fine-tuning on 34 paired examples from the phantom-like dataset.}
  \label{fig:realistic_generalization}
\end{figure}

To evaluate generalization under distribution shift, we simulate a more realistic resolution phantom-like dataset with structures and echo characteristics that much differ from the training data. Specifically, the new dataset consists of a speckle-dominated background formed by densely distributed weak scatterers, with additional structures including sparse point targets, hyperechoic regions with high reflectivity, and anechoic regions with no backscattered signal, thereby introducing structural and contrast variations beyond those seen during training.

Figure~\ref{fig:realistic_generalization}  presents the corresponding results. In particular, we evaluate two strategies on this dataset: direct zero-shot prediction and few-shot adaptation using only 34 paired examples. Overall, the model produces reasonably well beamformed images, with some expected local mischaracterization. Compared with direct beamforming of the $\times6$ sparse input (Fig.~\ref{fig:realistic_generalization}b), the zero-shot prediction (Fig.~\ref{fig:realistic_generalization}c) reduces structured sampling artifacts and suppresses prominent grating-lobe patterns, indicating nontrivial transfer beyond the training data. With few-shot adaptation~(Fig.~\ref{fig:realistic_generalization}d), the model further improves contrast consistency and speckle appearance while attenuating remaining artifacts. 

Overall, these results suggest that the proposed physics-guided supervision yields artifact-suppression behavior that can partially transfer across datasets under domain shift, and that limited fine-tuning can further narrow the gap to the fully sampled reference. That said, the outputs do not yet perfectly match the reference: residual intensity bias, texture mismatch, and localized artifacts remain, which we plan to address in future work.

\section{Conclusion and Future Work}

We develop a new physics-guided framework for reconstructing dense ultrasound RF data from sparse acquisitions. By coupling an RF-domain reconstruction objective with a beamforming-domain objective through a differentiable beamformer, the proposed method aligns RF interpolation with downstream image formation and improves artifact suppression under coherent DAS. To stabilize optimization across varying sampling conditions, we further introduced EMA-based adaptive loss balancing and random-skip masking during training to improve robustness to diverse decimation patterns. Experiments on simulated data showed that these components provide complementary benefits when combined. In particular, the best-performing configuration, which combines beamforming-guided supervision, EMA-based weighting, random-skip training, and a hybrid convolution--attention backbone, maintained mean SSIM near 0.95 across decimation factors from $\times2$ to $\times13$. These results indicate strong robustness under severe undersampling and encouraging generalization beyond the sparsity range seen during training.

Despite these promising results, several directions remain important in future works. Our study is based on simulated resolution phantom data, and validating the method on more realistic, patient-like datasets is a key next step. We plan to expand our datasets for further training and parameter fine-tuning. This will include rotationally-acquired linear array and convex array datasets imaging resolution phantoms and in vivo models. In the long-term, we aim to tailor these datasets towards longitudinal wearable ultrasound applications, including abdominal and fetal monitoring. With respect to further algorithm improvements, it would be valuable to explore self-supervised or semi-supervised objectives to reduce reliance on paired dense RF, extend the framework to 3D and multi-angle acquisitions, and profile latency and memory to support real-time deployment. These efforts should further narrow the gap between sparse acquisition constraints and high-quality ultrasound imaging in practical systems. Additionally, the model demonstrates a weakness in preserving anechoic regions. Because these structures are characterized by an absence of backscattered signal, rather than by strong, spatially coherent signals that are more readily interpolated, this limitation is not unexpected. However, it presents an important challenge for clinically relevant imaging targets such as the fluid-filled bladder, which typically appears anechoic on ultrasound. As a result, the current model's ability to generalize across common anatomical structures remains limited, constraining its present clinical utility. We aim to address this limitation in future work.

Overall, our results suggest that physics-guided RF interpolation is a promising direction for bridging sparse acquisition constraints and high-quality ultrasound imaging, with potential relevance to portable and wearable imaging systems.



\section*{Conflicts of Interest}
The authors declare that they have no conflicts of interest.
\vspace{0.5\baselineskip}

\section*{Author Contributions}
\noindent\textbf{Luoyuan Zhang}: Methodology, Investigation (experiments), Formal analysis, Visualization, Writing — original draft, Writing — review \& editing.\\
\textbf{Yinan Feng}: Methodology, Formal analysis, Writing — review \& editing.\\
\textbf{Ananya Tandri}: Data curation, Resources, Formal analysis, Investigation (experiments), Validation, Writing — review \& editing.\\
\textbf{Yiyang You}: Data curation, Formal analysis, Investigation (experiments), Validation, Writing — review \& editing.\\
\textbf{Hyunwoo Song}: Resources, Data curation, Writing — review \& editing.\\
\textbf{Jeeun Kang}: Conceptualization, Methodology,  Writing — review \& editing.\\
\textbf{Youzuo Lin}: Conceptualization, Supervision, Project administration,  Methodology, Writing — review \& editing.

\vspace{0.5\baselineskip}

\section*{Funding}

This work was supported by the University of North Carolina at Chapel Hill School of Data and Information Sciences through a faculty start-up grant, and by the U.S. National Science Foundation under Award No. 2504439. This work was supported by the National Institutes of Health Blueprint for Neuroscience Research (Grant Nos. U54EB015408 and U54EB033650).

\vspace{0.5\baselineskip}

\section*{Data Availability}
The data supporting the findings of this study are available from the corresponding author upon reasonable request. For peer review, the dataset used in this study is available at \url{https://drive.google.com/file/d/1BJCoot94ZSIpCbL7cAzNEV1QNUOvU-SH/view?usp=sharing}.
\vspace{0.5\baselineskip}

\section*{Acknowledgments}
\vspace{0.5\baselineskip}

This work was supported by the University of North Carolina at Chapel Hill School of Data Science and Society through a faculty start-up grant, and by the U.S. National Science Foundation under Award No. 2504439. This work was supported by the National Institutes of Health Blueprint for Neuroscience Research (Grant Nos. U54EB015408 and U54EB033650). Computational resources were provided by the University of North Carolina at Chapel Hill Information Technology Services Research Computing and Johns Hopkins University Advanced Research Computing at Hopkins (ARCH) core facility  (rockfish.jhu.edu) supported by the National Science Foundation (NSF) grant number OAC 1920103. 

\printbibliography


\renewcommand\theequation{\Alph{section}\arabic{equation}} 
\counterwithin*{equation}{section} 
\renewcommand\thefigure{\Alph{section}\arabic{figure}} 
\counterwithin*{figure}{section} 
\renewcommand\thetable{\Alph{section}\arabic{table}} 
\counterwithin*{table}{section} 







\end{document}